# Kappa-tail technique: Modeling and application to Solar Energetic Particles observed by Parker Solar Probe

G. Livadiotis[1], A.T. Cummings[1], M.E. Cuesta[1], R. Bandyopadhyay[1], H.A. Farooki[1], L.Y. Khoo[1], D.J. McComas[1], J.S. Rankin[1], T. Sharma[1], M.M. Shen[1], C.M.S. Cohen[2], G.D. Muro[2], and Z. Xu[2]

[1]*Department of Astrophysical Sciences, Princeton University, Princeton, NJ 08544, USA*
[2]*California Institute of Technology, Pasadena, CA, 91125, USA*

## Abstract

We develop the kappa-tail fitting technique, which analyzes observations of power-law tails of distributions and energy-flux spectra and connects them to theoretical modeling of kappa distributions, to determine the thermodynamics of the examined space plasma. In particular, we (i) construct the associated mathematical formulation, (ii) prove its decisive lead for determining whether the observed power-law is associated with kappa distributions; and (iii) provide a validation of the technique using pseudo-observations of typical input plasma parameters. Then, we apply this technique to a case-study by determining the thermodynamics of solar energetic particle (SEP) protons, for a SEP event observed on April 17, 2021, by the PSP/IS☉IS instrument suite onboard PSP. The results show SEP temperatures and densities of the order of ~1 MeV and ~$5 \cdot 10^{-7}$ cm$^{-3}$, respectively.

## 1. Introduction

NASA's Parker Solar Probe (PSP) mission is providing unique and unparalleled observations of solar energetic particle (SEP) events close in to their solar sources. Studies of these observations (e.g., McComas et al. 2019; Kouloumvakos et al. 2019; Wiedenbeck et al. 2020; Schwadron et al. 2020; Mitchell et al. 2020; Joyce et al. 2020; Desai et al. 2020; Cohen et al. 2021a; Giacalone et al. 2021; Dresing et al. 2023; Palmerio et al. 2024; Khoo et al. 2024; Cohen et al. 2024) have examined the time series of SEP fluxes over a broad range of energies and ions species (as well as electrons). Here we develop a different approach by characterizing the ensemble properties and thermodynamics of SEPs. This thermodynamic approach may provide deeper insight into the acceleration mechanisms that energize particles in the inner heliosphere and how the energetic particle distributions evolve and transport through the solar wind flow. This paper shows a case-study of our novel kappa-tail fitting technique, which analyzes observations of power-law tails of energy-flux spectra and connects them to the theoretical modeling of kappa distributions. We apply this technique to observations of SEPs to fit the kappa distribution function and show this leads to their thermodynamic parameters. Our companion paper by Cuesta et al. (2024) is a systematic application of this technique to SEP protons observations to characterize their overall thermodynamics.

Space plasmas throughout the heliosphere are characterized by correlations among the plasma particle velocities, which are induced by long-range interactions. These plasmas cannot be described within the classical framework of statistical mechanics and thermodynamics, which is based on the Boltzmann-Gibbs entropy and Maxwell-Boltzmann distribution of velocities (or energies) (Boltzmann 1866; Maxwell 1860).

Rather, space plasmas must be described by the generalized framework of nonextensive statistical mechanics and thermodynamics, all connected with the theoretical basis of kappa distributions (Livadiotis & McComas 2009, 2013; Livadiotis 2015a; 2017; Yoon 2019; Tsallis 2023).

The kappa distribution function describes the velocities (or energies) of plasma particle populations. It replaces the Maxwell-Boltzmann distribution for space plasmas and other complex particle systems, when correlations exist among their particles (Abe 1999; Livadiotis & McComas 2011a; Livadiotis 2018a;b; Livadiotis et al. 2021). The formulation of kappa distributions comprises the most generalized distribution function of velocities for systems residing in stationary states, which is called generalized thermal equilibrium (Livadiotis & McComas 2022; 2024). This is in contrast to classical thermal equilibrium, a single, extreme stationary state, where no particle correlations exist, i.e., particles move independent to each other, and as a result, only the Maxwell-Boltzmann distribution is allowed to characterize particles (e.g., Gibbs 1902; Livadiotis & McComas 2011a).

Kappa distributions have strong foundations within the context of (i) nonextensive statistical mechanics (Treumann 1997; Milovanov & Zelenyi 2000; Leubner 2002; Livadiotis & McComas 2009; Livadiotis 2014a, 2017, 2018a); (ii) thermodynamics (Livadiotis & McComas 2010; 2021; 2022; 2023a); and (iii) polytropic behavior (Livadiotis 2016; 2019; Nicolaou & Livadiotis 2019). Consequently, the kappa distribution uniquely provides the consistent characterization of temperature for systems residing in stationary states out of the classical thermal equilibrium (Livadiotis & McComas 2009, 2010, 2011a, 2021; Livadiotis 2009, 2017; Gravanis et al. 2020). The kappa distribution is characterized by temperature and kappa, the two independent parameters that determine the thermodynamic state of the particle system (Livadiotis 2018a). The full phase-space distribution, as well as the energy-flux spectrum (that is, the energy distribution of the differential flux), depends on the thermodynamic parameters of temperature and kappa, and also on the number density. While the temperature and density are well known plasma moments, the physical meaning of kappa within the framework of thermodynamics has just been recently understood (Livadiotis & McComas 2021; 2022; 2023a;b;c). This is interwoven solely with the correlations among particles; in fact, the correlation $\rho$ between the kinetic energies of any two particles per half degrees of freedom ($\frac{1}{2}D$), is exactly the inverse kappa, $\rho/(\frac{1}{2}D) = \frac{1}{\kappa}$; a formulation similar to the mean kinetic energy per half degrees of freedom, which is exactly the temperature, $<\varepsilon>/(\frac{1}{2}D) = k_\mathrm{B} T$. Lastly, we note that the thermodynamic kappa depends on the degrees of freedom $D$, $\kappa(D) = \kappa_0 + \frac{1}{2}D$, where the 0-degree kappa index, $\kappa_0$, is an invariant under variations of the dimensionality (Livadiotis & McComas 2011a; Livadiotis 2015b). Throughout the paper, we will be using any of the two notions of kappa, $\kappa_0$ or $\kappa = \kappa_0 + \frac{3}{2}$ (for $D=3$; see Section 3.1 for more discussions on dimensionality).

Moreover, kappa distributions have received impetus, as their "core" and "tail" features provide efficient modeling for observed distributions. In fact, due to their unique connection to thermodynamics, kappa distributions have been used to determine the temperature and other key thermodynamic parameters of particle populations in space plasmas throughout the heliosphere (e.g., see Livadiotis 2015a;c; 2017; Livadiotis et al. 2022). (In regards to solar wind particles, some examples are the following: Maksimovic et al. 1997; Pierrard et al. 1999; Mann et al. 2002; Marsch 2006; Zouganelis 2008, Livadiotis & McComas 2010; Yoon et al. 2012; Pavlos et al. 2016; Nicolaou & Livadiotis 2016; Livadiotis 2018c;d; Livadiotis et al. 2018; Wilson et al. 2019; Nicolaou et al. 2019; Silveira et al. 2021; Benetti et al. 2023.) There are a number of mechanisms generating kappa distributions in space plasmas, such as, superstatistics (e.g., Beck & Cohen 2003; Schwadron et al. 2010; Livadiotis et al. 2016; Gravanis et al. 2020; Davis et al. 2023; Ourabah 2024), effect of shock waves (e.g., Zank et al. 2006), turbulence (e.g., Yoon 2014; Bian et al. 2014), acceleration mechanism (e.g., Fisk & Gloeckler 2014), clustering (Peterson et al. 2013, Livadiotis et al. 2018); however, the transport of kappa distributions along the heliosphere is mainly affected by the pickup ions (Livadiotis & McComas 2011b; 2023b; 2024).

Despite the tremendous applications of kappa distributions in understanding the thermodynamics of space plasma particles, the thermodynamics of SEPs have only been rarely examined (e.g., Xiao et al. 2008, Laming et al. 2013, Lee et al. 2024), and to date, have never been methodologically derived in association with kappa distributions.

SEPs have been observed since the middle of the 20th century (e.g., Forbush 1946). They represent a significant aspect of space weather phenomena, and are understood to be accelerated through a variety of phenomena, including solar flares and shocks, often driven by coronal mass ejections (CMEs) in the inner heliosphere and corotating stream interactions farther out in the heliosphere (Fisk 1971; Axford et al. 1977; Blandford & Ostriker 1978; Bell 1978; Gosling 1979; Lee 1983; Kallenrode 1996; Lee 1997; Lario et al. 1998). The thermodynamic properties of SEPs could provide a new method for understanding, accurately modeling, and even predicting their behavior. In this study, we examine temporal profiles of the thermodynamic quantities of SEPs observed by the Integrated Science Investigation of the Sun (IS☉IS) instrument (McComas et al. 2016) onboard PSP (Fox et al. 2016). By modeling the observed energy-flux spectrum with the formalism of kappa distributions we derive the SEP's thermodynamic parameters of temperature, density, and kappa.

Observations of the energy ($\varepsilon$) – flux ($J$) spectrum for energetic particles, typically refer to its tail, which is characterized by a power-law behavior. Then, a linear relationship is used for plotting and analyzing the spectra on log-log scales, (log $J$ versus log $\varepsilon$). A power-law distribution has the same behavior (linear, on log-log scales) throughout the spectrum; otherwise, an asymptotic power-law tail distribution exhibits concave or convex behavior (i.e., positive or negative curvature) to the extent of the spectrum away from the power-law tail; (also see the context of heavy-tail distributions, e.g., Asmussen 2003). Kappa distributions are employed as the model function for describing and fitting the energy-flux spectrum. For example, the kappa distributions are characterized by a concave plot away from the tail and near the core, while the presence of pickup ions may lead to convex distributions; (e.g., Livadiotis et al. 2012). Even though kappa distributions have fundamental importance, with their origin uniquely connected to thermodynamics, other mathematical formulations may describe convex or concave distributions with power-law tails (see Figure 1). Thus, the question arises of how we can test the observed power-law tail of a spectrum to determine whether it is connected with a kappa distribution.

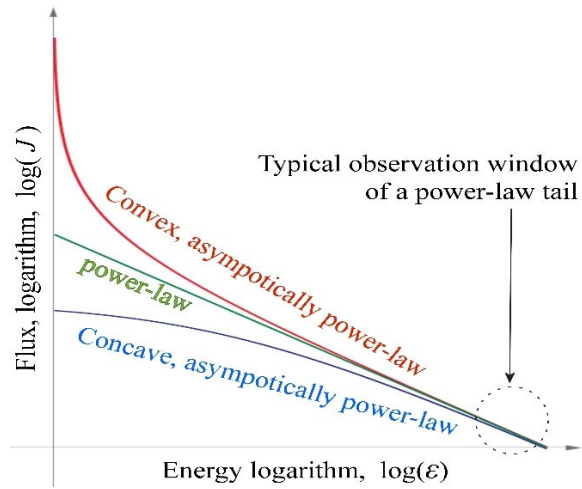

**Figure 1.** Asymptotically power-law tails can be connected to concave or convex energy-flux spectra. The kappa-tail fitting technique determines whether the observed power-law tail is associated with kappa distributions.

For this, we develop a novel technique that determines whether the power-law tails are associated with kappa distributions. In addition, this technique is used for deriving the thermodynamic parameters of temperature and density. This is based on two levels of linear fits. First, the linear fit of the energy-flux spectrum on log-log scales leads to the estimation of the spectral slope and intercept of the linear tail; (the absolute value of the slope is also known as the spectral index). These two parameters have a special linear relationship when the spectrum is connected with kappa distributions. The analysis of several spectra, sequential in time, leads to the derivation of several pair values of spectral index and intercept, whose linear relationship is a signature of kappa distributions. Then, the $2^{nd}$ level of fitting involves modeling this linearity between the spectral index and intercept, where the new slope and intercept, i.e., the resultant of the $2^{nd}$ fit, are respectively related to the temperature and density of the examined particles, in this case, SEPs.

The purpose of this paper is to derive the fundamentals for kappa-tail fitting technique and show a case study for analyzing and interpreting the energy-flux spectra of SEPs in association with kappa distributions modeling. This procedure determines the temperature and other thermodynamic parameters, characterizes the thermodynamic state of the observed energetic particles, and finally allows us to compare their thermodynamics with that of the solar wind plasma particles. The proof-of-concept of the novel technique associated with kappa distributions is shown by examining the thermodynamic properties of SEPs for two arbitrary intervals during a SEP event associated with CME observed by PSP on April 17, 2021 (Dresing et al. 2023).

The paper is organized as follows. In Section 2, we present the theoretical model of the power-law spectra originated from kappa distributions. In Section 3, we show the novel kappa-tail fitting technique used for determining (a) whether the tail is associated with kappa distributions, and (b) the temperature and density of the examined particles, here, SEPs. In particular, we (i) develop the mathematical formulation of the technique, (ii) prove that the technique can be used for determining a "signature" of kappa distributions, which is decisive on whether the observed power-law is associated with these distributions; and (iii) provide a validation of the technique using pseudo-observations of typical input plasma parameters. In Section 4, we apply the kappa-tail fitting technique to observed SEPs. Specifically, we (i) explain the data product from PSP and the application of the technique with which we assess thermodynamic properties of SEP protons observed by PSP; (ii) apply this technique to two separate, arbitrary intervals of a SEP event, and (iii) estimate the temperature and density of high energy protons for each corresponding interval. Finally, in Section 5, we discuss the results and conclusions of this study, while the two appendices show parts of the technique in more details.

**2. Theoretical Model: Power-law spectra originated from kappa distributions**

We derive the expressions used for describing the spectrum of the proton differential flux with energy, based on the kappa distribution theory and formalism. The 3D distribution of the velocities of plasma particles is described by a kappa distribution (Livadiotis & McComas 2013; Livadiotis 2017). Here, we focus on the modeling of energetic protons; nevertheless, this distribution function constitutes the appropriate formula for describing the thermodynamics of any particle populations residing in stationary states (Livadiotis 2018a; Livadiotis & McComas 2022, 2023a;b, 2024); this is given by:

$$P_u(\boldsymbol{u}) = \pi^{-\frac{3}{2}} \cdot N(\kappa_0) \cdot \theta^{-3} \cdot \left[1 + \frac{1}{\kappa_0} \frac{(\boldsymbol{u} - \boldsymbol{u}_b)^2}{\theta^2}\right]^{-\kappa_0 - \frac{5}{2}}. \qquad (1)$$

which involves the particle $\boldsymbol{u}$ and bulk $\boldsymbol{u}_b$ velocities of the plasma particles. In this formulation, the temperature is expressed in terms of the effective speed-scale parameter $\theta = \sqrt{2k_B T / m}$, that is, the temperature in speed dimensions, where $k_B$ is the Boltzmann's constant, and $m$, $T$ are the particle mass and temperature, respectively; the thermodynamic kappa parameter is given by $\kappa = \kappa_0 + \frac{3}{2}$ (for $D=3$), while the normalization coefficient, $N(\kappa_0)$, is a pure function of kappa,

$$\mathrm{N}(\kappa_0) \equiv \kappa_0^{-\frac{3}{2}} \cdot \frac{\Gamma(\kappa_0 + \frac{5}{2})}{\Gamma(\kappa_0 + 1)} \;, \tag{2}$$

with $\mathrm{N}(\kappa_0 \to \infty) = 1$ (at the classical thermal equilibrium).

In terms of the kinetic energy, measured in the comoving frame, $\varepsilon = \tfrac{1}{2} m \cdot (\boldsymbol{u} - \boldsymbol{u}_\mathrm{b})^2$, the distribution (1) becomes

$$P_u(\varepsilon) = \pi^{-\frac{3}{2}} \cdot \mathrm{N}(\kappa_0) \cdot \left(\frac{2 k_\mathrm{B} T}{m}\right)^{-3/2} \cdot \left(1 + \frac{1}{\kappa_0} \cdot \frac{\varepsilon}{k_\mathrm{B} T}\right)^{-\kappa_0 - \frac{5}{2}}, \tag{3a}$$

or, its logarithm,

$$\log P_u(\varepsilon) = \log\left[\pi^{-\frac{3}{2}} \mathrm{N}(\kappa_0)(2 k_\mathrm{B} T / m)^{-\frac{3}{2}}\right] - (\kappa_0 + \tfrac{5}{2}) \cdot \log\left(1 + \frac{1}{\kappa_0} \cdot \frac{\varepsilon}{k_\mathrm{B} T}\right). \tag{3b}$$

The kappa distribution is reduced into a power-law relationship when $\varepsilon \gg \kappa_0 k_\mathrm{B} T$, leading to its characteristic heavy tail, i.e., $P_u(\varepsilon) \propto \varepsilon^{-\kappa_0 - \frac{5}{2}}$. Indeed, expanding Eq.(3b) up to the parabolic term, we obtain

$$\begin{aligned}
\log\left(1 + \frac{1}{\kappa_0} \cdot \frac{\varepsilon}{k_\mathrm{B} T}\right) &= \log\left(\frac{\varepsilon}{\kappa_0 k_\mathrm{B} T}\right) + \log e \cdot \ln\left(1 + \frac{\kappa_0 k_\mathrm{B} T}{\varepsilon}\right) \\
&\cong \log(\varepsilon / [\mathrm{keV}]) - \log(\kappa_0 k_\mathrm{B} T / [\mathrm{keV}]) + \frac{\log e \cdot \kappa_0 k_\mathrm{B} T}{\varepsilon} + O\left(\frac{\kappa_0 k_\mathrm{B} T}{\varepsilon}\right)^2.
\end{aligned} \tag{4}$$

Hence, having Eq.(4) substituted into Eq.(3b), the distribution becomes

$$\begin{aligned}
\log P_u(\varepsilon) &\cong \log\left[\pi^{-\frac{3}{2}} \mathrm{N}(\kappa_0)(2 k_\mathrm{B} T / m)^{-\frac{3}{2}}\right] + (\kappa_0 + \tfrac{5}{2}) \cdot \log(\kappa_0 k_\mathrm{B} T / [\mathrm{keV}]) \\
&\quad - (\kappa_0 + \tfrac{5}{2}) \cdot \log\left(\frac{\varepsilon}{[\mathrm{keV}]}\right) - \frac{\log e \cdot \kappa_0 (\kappa_0 + \tfrac{5}{2}) \cdot k_\mathrm{B} T}{\varepsilon}.
\end{aligned} \tag{5}$$

Equation (5) comprises the linear expansion of the logarithm of the distribution function in terms of the logarithm of energy, while the additional nonlinear term becomes negligible at high energies. We verify the validity of this linear approximation: for the typical values of $\kappa_0 \sim 2.5$, $k_\mathrm{B} T \sim 0.8$ MeV, $\varepsilon \sim 10$ MeV characterizing the SEP population, we estimate the last two terms and find the ratio 1:33 of the nonlinear $\log e \cdot \kappa_0 (\kappa_0 + \tfrac{5}{2}) \cdot (k_\mathrm{B} T / \varepsilon)$ compared to the linear $(\kappa_0 + \tfrac{5}{2}) \cdot \log(\varepsilon / [\mathrm{keV}])$ terms.

Next, we consider the particle flux, expressed in terms of the distribution, that is,

$$J(\vec{u}) = \frac{1}{m} \cdot n \cdot P_u(\vec{u}) \cdot u^2 \;\; \text{or} \;\; J(\varepsilon) = 2 m^{-2} \cdot n \cdot P_u(\varepsilon) \cdot \varepsilon \;. \tag{6}$$

Setting the units, the flux logarithm becomes:

$$\begin{aligned}
\log\left(\frac{J}{[\mathrm{cm}^{-2}\,\mathrm{s}^{-1}\,\mathrm{sr}^{-1}\,\mathrm{keV}^{-1}]}\right) &\cong \log\left[2^{-\frac{1}{2}} \pi^{-\frac{3}{2}} \left(\frac{m}{[\mathrm{kg}]}\right)^{-\frac{1}{2}} \left(\frac{[\mathrm{keV}]}{[\mathrm{j}]}\right)^{\frac{1}{2}} \frac{[\mathrm{m}]}{[\mathrm{cm}]}\right] \\
&\quad + \log\left(\frac{n}{\mathrm{cm}^{-3}}\right) + (\kappa_0 + 1) \cdot \log\left(\frac{k_\mathrm{B} T}{[\mathrm{keV}]}\right) + \log\left[N(\kappa_0) \cdot \kappa_0^{\kappa_0 + \frac{5}{2}}\right] \\
&\quad - (\kappa_0 + \tfrac{3}{2}) \cdot \log\left(\frac{\varepsilon}{[\mathrm{keV}]}\right) - \log e \cdot \kappa_0 (\kappa_0 + \tfrac{5}{2}) \cdot \frac{k_\mathrm{B} T}{[\mathrm{keV}]} 10^{-\log(\varepsilon / [\mathrm{keV}])}.
\end{aligned} \tag{7}$$

This can be written in the form of a statistical model, i.e.,

$$Y^{(1)} \cong \log \text{Int}^{(1)} + \text{Slo}^{(1)} \cdot X^{(1)} + \text{NLin}^{(1)} \cdot 10^{-X^{(1)}} \ , \tag{8a}$$

where we set the variables:

$$X^{(1)} \equiv \log\left(\frac{\varepsilon}{[\text{keV}]}\right) \ , \ Y^{(1)} \equiv \log\left(\frac{J}{[\text{cm}^{-2}\ \text{s}^{-1}\ \text{sr}^{-1}\ \text{keV}^{-1}]}\right), \tag{8b}$$

and the fitting parameters:

$$\log \text{Int}^{(1)} \equiv C + \log\left(\frac{n}{\text{cm}^{-3}}\right) + (\kappa_0 + 1) \cdot \log\left(\frac{k_B T}{\text{keV}}\right) + \log\left[\kappa_0^{\kappa_0+1} \cdot \frac{\Gamma(\kappa_0 + \frac{5}{2})}{\Gamma(\kappa_0 + 1)}\right],$$
$$\text{Slo}^{(1)} \equiv -(\kappa_0 + \tfrac{3}{2}), \tag{8c}$$
$$\text{NLin}^{(1)} \equiv -\log e \cdot \kappa_0 (\kappa_0 + \tfrac{5}{2}) \cdot \frac{k_B T}{[\text{keV}]}.$$

The involved constant is given by:

$$C \equiv \log\left[2^{-\frac{1}{2}}\pi^{-\frac{3}{2}}\left(\frac{m}{[\text{kg}]}\right)^{-\frac{1}{2}}\left(\frac{[\text{keV}]}{[\text{j}]}\right)^{\frac{1}{2}}\frac{[\text{m}]}{[\text{cm}]}\right] = \log\left(\frac{100}{\sqrt{2\pi^3 m [\text{keV}]}}\right) \cong 6.60 \ . \tag{8d}$$

It is clear that the power-law tail of kappa distributed energy-flux spectra, exhibiting a linear relationship between energy and flux when plotted on logarithmic scales, occurs at sufficient high energies, $\varepsilon \gg \kappa_0 k_B T$, where we can ignore the last nonlinear term involved in Eq.(8c), which is proportional to $\kappa_0 k_B T / \varepsilon$.

## 3. Methodology
### 3.1. The kappa-tail fitting technique

We develop the technique for fitting the power-law tail of kappa distributions. The technique consists of two levels of fitting conducted on the tail of the energy-flux spectrum. In the 1st level of fitting, the log-log plot of the energy-flux spectrum is fitted with a linear statistical model, according to

$$Y^{(1)} \cong \log \text{Int}^{(1)} + \text{Slo}^{(1)} \cdot X^{(1)} \ , \tag{9}$$

where we seek to find the two fitting parameters, the intercept $\log \text{Int}^{(1)}$ and the slope $\text{Slo}^{(1)}$.

Then, we derive the spectral index, $\gamma$, and the modified intercept, $\log \text{Int}'$: The spectral index is typically defined by the absolute value of the slope of the energy-flux spectrum, expressed on a log-log scale, and coincides with the thermodynamic kappa; on the other hand, the modified intercept is defined accordingly to enclose the nonlinear terms of the thermodynamic kappa; namely,

$$\gamma \equiv -\text{Slo}^{(1)} = \kappa_0 + \tfrac{3}{2} \ , \tag{10a}$$

and

$$\log \text{Int}' \equiv \log \text{Int}^{(1)} - \log\left[\kappa_0^{\kappa_0+1} \cdot \frac{\Gamma(\kappa_0 + \frac{5}{2})}{\Gamma(\kappa_0 + 1)}\right]$$
$$\cong C + \log\left(\frac{n}{\text{cm}^{-3}}\right) + (\kappa_0 + 1) \cdot \log\left(\frac{k_B T}{\text{keV}}\right). \tag{10b}$$

(Note that the approximately equal sign comes from the expansion of the flux logarithm in Eq.(7)). Then, we define a new set of variables $(X^{(2)}, Y^{(2)})$, as follows:

$$X^{(2)} \equiv \gamma - \tfrac{1}{2} , \quad Y^{(2)} \equiv \log \text{Int}' = \log \text{Int}^{(1)} - \log[A(\gamma)] , \quad (11)$$

where we set the characteristic function $A(\gamma)$ that modifies the intercept

$$A(\gamma) \equiv (\gamma - \tfrac{3}{2})^{\gamma+1} N(\gamma - \tfrac{3}{2}) = (\gamma - \tfrac{3}{2})^{\gamma - \tfrac{1}{2}} \cdot \frac{\Gamma(\gamma+1)}{\Gamma(\gamma - \tfrac{1}{2})} . \quad (12)$$

Then, Eq.(10b) is written as

$$Y^{(2)} \cong C + \log\left(\frac{n}{[\text{cm}^{-3}]}\right) + \log\left(\frac{k_B T}{[\text{keV}]}\right) \cdot X^{(2)} , \quad (13)$$

where we observe that various measurements of the pair values $(X^{(2)}, Y^{(2)})$ could potentially lead to an estimation of the values of temperature and density. This is the task of the 2nd level of fitting, but before doing that we need a set of aggregated pair values of $(X^{(2)}, Y^{(2)})$.

The basic property of $X^{(2)}$ and $Y^{(2)}$ is that both depend on the polytropic process, and thus, the variation of the polytropic parameters corresponds to a variation of the values of $(X^{(2)}, Y^{(2)})$. The spectral index and the modified intercept remain invariant along a streamline of the flow during a polytropic process. This is because the spectral index $\gamma$ is related to the polytropic index $\nu$, i.e., $\nu + \gamma = \tfrac{1}{2}$, while the modified intercept is related to the polytropic pressure $\Pi = const. \cdot n \cdot T^{-\nu}$ (e.g., see: Livadiotis 2016; 2019; Livadiotis et al. 2022). Both the polytropic parameters $\nu$ and $\Pi$ are invariant along a polytropic process. Slow variation of these parameters leads to a similar variation of the spectral index and the modified intercept, that is, of the pair values of $(X^{(2)}, Y^{(2)})$. On the other hand, sequential spectra are characterized by similar thermodynamics, and thus, they are parameterized by close values of temperature and density. Thus, the values of the modified intercept and spectral index vary, constructing different pair values of $(X^{(2)}, Y^{(2)})$, while the involved temperature and density are treated as fixed parameters to be determined by the linear fitting of $(X^{(2)}, Y^{(2)})$.

The set of aggregated pair values of $\{X^{(2)}, Y^{(2)}\}$, needed for the 2nd level of fitting, is constructed as follows. We assume that the parameters of the polytropic process characterizing the energetic protons have slow variation along the time series interval at which we apply the technique. (This can be tested by examining the goodness of the fits involved in the technique.) Thus, we separate the whole-time interval $\Delta t$ in smaller subintervals, $\Delta t_i$, with $\Delta t = (\Delta t_1 + \Delta t_2 + \ldots + \Delta t_N)$, each one determining a slightly different pair of values, $\{(X_i^{(2)}, Y_i^{(2)})\}_{i=1}^{N} = (X_1^{(2)}, Y_1^{(2)}), (X_2^{(2)}, Y_2^{(2)}), \cdots, (X_N^{(2)}, Y_N^{(2)})$. The 2nd level of (linear) fitting involves using these fluctuated pair values. The expectation is that the plot of $(X^{(2)}, Y^{(2)})$ values exhibits also a linear behavior, given by Eq.(13). For this, we examine a number of $N$ spectra sequential in time. Each spectrum fit provides a pair values of $(X^{(2)}, Y^{(2)})$, so from all the sequential spectra we collect a dataset $\{(X_i^{(2)}, Y_i^{(2)})\}_{i=1}^{N}$, at which we fit the linear model:

$$Y^{(2)} \cong \log\text{Int}^{(2)} + \text{Slo}^{(2)} \cdot X^{(2)} . \qquad (14)$$

This linear relationship between the spectral index ($X^{(2)}$) and modified intercept ($Y^{(2)}$) stands as a signature of the presence of kappa distributions (Livadiotis et al. 2011; 2022); this is because the linearity holds only when the values of $Y^{(2)}$ are extracted using the characteristic function A given by Eq.(12). In addition, the new parameters of intercept $\log\text{Int}^{(2)}$ and slope $\text{Slo}^{(2)}$ can provide the density and temperature of this population, respectively; namely:

$$\log\text{Int}^{(2)} \equiv C + \log\left(\frac{n}{[\text{cm}^{-3}]}\right) , \quad \text{Slo}^{(2)} \equiv \log\left(\frac{k_B T}{\text{keV}}\right) , \qquad (15a)$$

or, solving in terms of the thermodynamic parameters,

$$n = 2.51 \times 10^{-7} \times \text{Int}^{(2)} [\text{cm}^{-3}] , \quad T = 10^{\text{Slo}^{(2)}} [\text{keV}] , \qquad (15b)$$

where we set $\text{Int}^{(2)} = 10^{\log\text{Int}^{(2)}}$.

Finally, one mechanism that can cause the variability of the polytropic processes is the effective dimensionality. There are several causes that decrease the effective dimensionality of protons $D$ of the standard value of the embedded 3-dimensional space ($D$=3). This can be (1) the anisotropy of the distribution caused by the magnetic field (e.g., Livadiotis & Nicolaou 2021; Livadiotis & McComas 2023b; Livadiotis et al. 2024) leading to distributions closer to 2-dimensional (pie-type) or 1-dimensional (cigar-type), (2) the presence of newly born pick-up ions, orbiting the field in a 1-dimensional ring, merge with the plasma protons reducing its dimensionality (Livadiotis & McComas 2023b; Livadiotis et al. 2024). In this latter case, the $D$-dimensional distribution embedded in a 3-dimensional space is given again by Eq.(3), but now the kappa-like parameter involved in the distribution is $\kappa_{0,D} \equiv \kappa_0 + \frac{D-3}{2}$. Therefore, the previous technique for finding the density and temperature can be repeated as already described. The only difference is that the variability of the spectral index would be caused by the variability of the effective dimensionality $D$, because of $\gamma = \kappa_{0,D} = \kappa_0 + \frac{D-3}{2}$ replacing Eq.(10a).

### 3.2. The "signature" of kappa distributions in the technique

If the velocity distribution, $P_u(\boldsymbol{u})$ or $P_u(\varepsilon)$, was not given by a kappa function, but instead, by an arbitrary function $f$, then, this would have been expressed as

$$P_u(\varepsilon) = \pi^{-\frac{3}{2}} \cdot N_f \cdot \left(\frac{2k_B T}{m}\right)^{-3/2} \cdot f\left(\frac{\varepsilon}{k_B T}\right) , \qquad (16)$$

and from Eq.(6), we derive the respective flux

$$J(\varepsilon) \propto \mathrm{N}_f \cdot n \cdot (k_\mathrm{B} T)^{-3/2} \cdot f\left(\frac{\varepsilon}{k_\mathrm{B} T}\right) \cdot \varepsilon, \tag{17}$$

where $\mathrm{N}_f$ is the normalization constant. Then, the flux of a power-law tail, $J(\varepsilon \gg k_\mathrm{B} T) \sim \varepsilon^{-\gamma}$, requires $f(x) \to f_\infty(\gamma) \cdot x^{-\gamma-1}$, or

$$J(\varepsilon \gg k_\mathrm{B} T) \propto \mathrm{N}_f(\gamma) \cdot f_\infty(\gamma) \cdot n \cdot (k_\mathrm{B} T)^{\gamma-1/2} \cdot \varepsilon^{-\gamma}, \tag{18}$$

from where we derive that the characteristic function is (similar to Eq.(12)):

$$\mathrm{A}_f(\gamma) \equiv \mathrm{N}_f(\gamma) \cdot f_\infty(\gamma). \tag{19}$$

Without any loss of its generality, we may write the function $f$, as an asymptotic power-law $f(x) = f_\infty(\gamma) \cdot x^{-\gamma-1} \cdot M(x)$, expressed in terms of a bounded function $M(x)$, i.e., $M(x \to \infty) = 1$. Then, the characteristic function is given by a Mellin transform of $M(x)$ (e.g., see Mellin 1897),

$$1/\mathrm{A}_f(\gamma) = \int_0^\infty M(x) \cdot x^{-\gamma-1/2} dx. \tag{20}$$

Given $\mathrm{A}_f(\gamma)$, the function $M(x)$, and thus, also the distribution function $f(x)$, can be exactly determined. Therefore, we derive that the characteristic function is one-to-one with the distribution function, namely, the characteristic function $\mathrm{A}_f$ is a signature of the distribution function $f$, and vice versa.

Next, from Eq.(18), we express the linear tail in log-log scales,

$$\log J(\varepsilon \gg k_\mathrm{B} T) = \log \mathrm{Int}^{(1)} - \gamma \cdot \log \varepsilon, \tag{21}$$

where the intercept is written as

$$\log \mathrm{Int}' \equiv \log \mathrm{Int}^{(1)} - \log\left[\mathrm{A}_f(\gamma)\right] = C_f + \log\left(\frac{n}{\mathrm{cm}^{-3}}\right) + (\gamma - 1/2) \cdot \log\left(\frac{k_\mathrm{B} T}{\mathrm{keV}}\right) \tag{22}$$

where $C_f$ is set to include all the constant values.

We observe that in order for the right-hand side of Eq.(22) to be linear with respect to spectral index, the intercept must be modified by subtracting the logarithm of the characteristic function $\mathrm{A}_f(\gamma)$. If we were subtracting the characteristic function of any other distribution, the resultant right-hand side won't be linear in terms of $\gamma$, due to the nonzero quantity $G(\gamma) \equiv \log\left[\mathrm{A}_f(\gamma)/\mathrm{A}_g(\gamma)\right]$,

$$\log \mathrm{Int}' \equiv \log \mathrm{Int}^{(1)} - \log\left[\mathrm{A}_g(\gamma)\right] = C_f + \log\left(\frac{n}{\mathrm{cm}^{-3}}\right) + (\gamma - 1/2) \cdot \log\left(\frac{k_\mathrm{B} T}{\mathrm{keV}}\right) + G(\gamma). \tag{23}$$

Therefore, if we use for statistical modeling a characteristic function $\mathrm{A}_g$, different than the actual characteristic function $\mathrm{A}_f$ embedded in the population distribution, then, the existence of the nonzero nonlinear function $G(\gamma)$ (in the right-hand side of Eq.(23)) would have significantly decreased the goodness of the linear fitting between modified intercept and spectral index ($2^\mathrm{nd}$ level fitting). This property can be used for testing whether we have selected the "right" characteristic function, since the choice of $\mathrm{A}_g = \mathrm{A}_f$ would have eliminated the nonlinear effects of the $2^\mathrm{nd}$ level fitting and maximized the fitting goodness. Consequently, if the actual characteristic function is the one associated with a kappa distribution, and we use the same characteristic function, as given by Eq.(12), for modeling the modified intercept, then, the 2$^{nd}$ level of fitting would have been linear. Practically, we use the characteristic function A($\gamma$) of a kappa distribution, as given by Eq.(12), and if we detect a linear relationship between spectral index $\gamma$ and modified intercept logInt′ (acceptable under some statistical confidence), then, we would conclude that the kappa distribution is the underlying distribution function associated with the observed power-law tail.

### 3.3. Validation of the technique

For the validation of the kappa-tail fitting technique, developed in the previous subsection, we produce pseudo-observations for typical input SEP parameters, and then, apply the two fitting levels of the technique.

Nine of the energy bands, characteristic of the IS☉IS instrument's EPI-Hi HETA (pointing in the quasi-sunward direction) onboard PSP, is shown in Figure 2(a), starting from $\approx$10 to 50 MeV (see: Sarlis et al. 2024 and references therein). Note that the event described in the application in Section 4 is weak and a non-zero signal is found only in those energy channels from 10-50 MeV, even though we are including signal from 10-80 MeV; here, for a consistent validation, we use the same energy bands.

We start with Eq.(7) for expressing the flux logarithm; then, we add noise on the flux values modelled by kappa distributions $J(\kappa)$ according to

$$\log J = \log J(\kappa) + \sigma \cdot r / \ln(100) , \qquad (24)$$

where $r$ is a random number in the interval [-1,1], and the disturbed flux value $J$ can be expressed in terms of the relative error $\sigma$. Figure 2(b) shows the energy-flux spectrum for three different relative errors $\sigma$ of added noise.

We recall that the linear relationship between the modified intercept of the flux logarithm, $Y^{(2)}$, and the spectral index, $X^{(2)}$, stands as a signature of the presence of kappa distributed velocities (Livadiotis et al. 2011; 2022). This linearity holds only when using the characteristic function A($\gamma$) to evaluate $Y^{(2)}$. Here, we verify that any alternative of kappa distributions has lower statistical confidence than the case of kappa distributions. We show this, by (i) setting a function A, alternative to the one shown in Eq.(12) by an exponent $\alpha$, that is,

$$A(\gamma;\alpha) \equiv \left[(\gamma-\tfrac{3}{2})^{\gamma-\tfrac{1}{2}} \cdot \frac{\Gamma(\gamma+1)}{\Gamma(\gamma-\tfrac{1}{2})}\right]^{\alpha} , \text{ or } \log[A(\gamma;\alpha)] = \alpha \cdot \log[A(\gamma;\alpha=1)] , \qquad (25)$$

and then, (ii) estimating the optimal value of $\alpha$ as the one that minimizes the respective chi-square. In particular, the chi-square is set by three fitting parameters, according to

$$\text{Chi-Square}(\alpha;\text{logInt}^{(2)},\text{Slo}^{(2)}) = \sum_{i=1}^{N} \left\{\frac{Y^{(2)}_i(\alpha) - [\text{logInt}^{(2)} + \text{Slo}^{(2)} \cdot X^{(2)}_i]}{(\sigma_{Y^{(2)}})_i}\right\}^2 . \qquad (26)$$

Figure 2(c) shows that the chi-square in Eq.(26) is minimized for $\alpha \sim 1$, for all values of $\sigma$. Also, in Figure 2(d), we observe the linearity between the pair values of $\{(X_i^{(2)}, Y_i^{(2)})\}_{i=1}^N$. (Note that deviations from the linearity may be observed for spectral indices close to the lowest limit of kappa, i.e., $\gamma \sim 1.5$, but the respective large uncertainties of these points do not have any impact on the statistics of the linearity.) Given this linear relationship, we determine the temperature and density from the respective slope and intercept, as shown in Eqs.(15a,b). These results are shown in Table 1, which are close (within statistical confidence) to the original values of $\log(T/[\text{keV}]) = -2.69$ and $\log(n/[\text{cm}^{-3}]) = -6$.

**Table 1. Derivation of temperature and density for pseudo - observations**

| $\sigma$ | $\log(T/[\text{keV}])$ | $\delta\log(T/[\text{keV}])$ | $\log(n/[\text{cm}^{-3}])$ | $\delta\log(n/[\text{cm}^{-3}])$ |
|---|---|---|---|---|
| 5% | 2.60 | 0.22 | -6.03 | 0.28 |
| 10% | 2.58 | 0.41 | -6.01 | 0.53 |
| 15% | 2.57 | 0.58 | -5.99 | 0.75 |

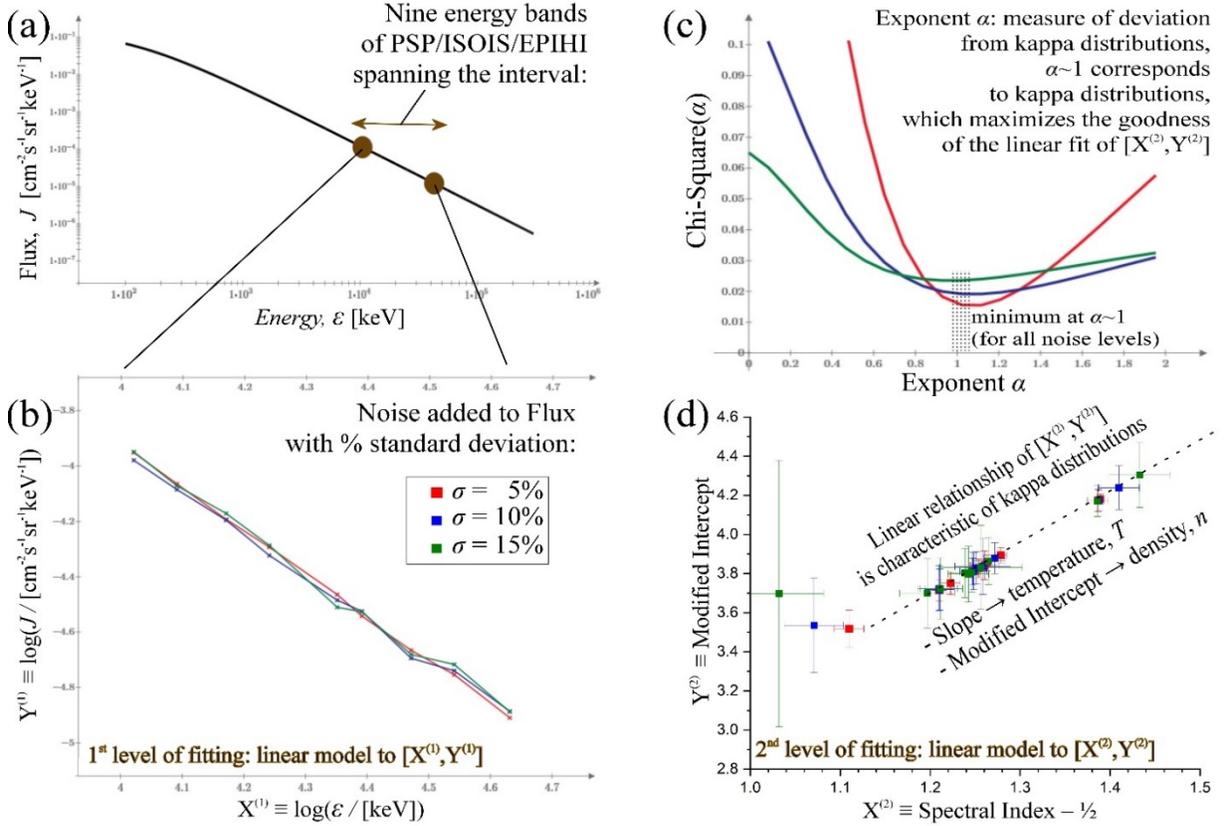

**Figure 2.** (a) Energy bands of the EPI-HI HETA onboard PSP/ IS☉IS. (b) Energy-flux kappa distributed spectrum for three different relative errors $\sigma$ of added noise (=5, 10, and 15%, represented by red, blue, and green line, respectively). (c) Chi-square, shown in Eq.(26), minimized for $\alpha \sim 1$, and for all different values of $\sigma$. (d) The linear relationship between the pair values of $\{(X_i^{(2)}, Y_i^{(2)})\}$, where we determine the temperature and density from the respective slope and intercept.

### 3.4. Variation of the technique for detecting nonstationary transitions

The technique consists of two levels of linear fitting conducted on the tail of the energy-flux spectrum. Especially the second level of fitting is decisive on the formulation of the velocity distribution function. Nevertheless, the stationarity of the examined timeseries is required. Namely, the sequential timeseries intervals, which produce the aggregated pairs of $\{X^{(2)}, Y^{(2)}\}$ to be used for the second level of linear fit, must be characterized by the same stationary state and velocity distribution function. On the contrary, when the examined series of time intervals are not characterized by a stationary state, but instead, exhibit a transition between different stationary states, or, a more complicated nonstationary evolution, then, the technique fails. In this case, we use a variation of the technique; that is, we solve the system of equations for finding the parameters, slope $\mathrm{Slo}^{(2)}$ and intercept $\log \mathrm{Int}^{(2)}$, over a subset of the sequential intervals, and then, average their values over the whole set of the sequential intervals. These estimated averages, $\overline{\log \mathrm{Int}^{(2)}}$ and $\overline{\mathrm{Slo}^{(2)}}$, can be used to determine whether the examined interval is stationary or is characterized by transition between stationary states.

For this, we compare the estimated average $\bar{p}$ with the respective value $p_{\mathrm{s.f.}}$ estimated from the second level fitting (where $p$ indicates any of the two parameters of slope $\mathrm{Slo}^{(1)}$ and intercept $\log \mathrm{Int}^{(1)}$). If there were no transitions, the respective values would have been equal (under a statistical confidence). We perform a $t$-test (e.g., Sawilowsky & Blair 1992) to confirm or reject this hypothesis; the value of $t_{\mathrm{est}}$ is determined by $t_{\mathrm{est}} = |\bar{p} - p_{\mathrm{s.f.}}| / \sqrt{\delta \bar{p}^2 + \delta p_{\mathrm{s.f.}}^2}$ and is used for finding the P-value, which measures the probability of having a value of $t$ larger than the estimated, $t_{\mathrm{est}}$. Then, we distinguish among the following cases: (i) When the estimated P-value is ≥0.05, the hypothesis of equal averaged $\bar{p}$ and fitted $p_{\mathrm{s.f.}}$ values is accepted. (ii) When the P-value is smaller, but ≥0.01, we accept that there is a weak transition in the examined time interval. In this case, the median should be used instead of the mean. (In Appendix A, we describe the variation of the technique, in detail. In Appendix B, we show the quality of median for describing the values of slope $\mathrm{Slo}^{(1)}$ and intercept $\log \mathrm{Int}^{(1)}$.) Finally, when the P-value is even smaller, i.e., < 0.01, we reject the hypothesis, and the examined time interval is noted as nonstationary.

### 4. Application to SEP events
### 4.1. Data

PSP continues to provide unprecedented proximity to the Sun on each successive orbit, enabling the collection of invaluable data on SEPs. PSP/IS☉IS measures the composition and properties of SEPs, offering insights into their energetic characteristics as summarized in the introduction.

Here we show how to process PSP data to extract the bulk values of SEP temperature and density from an SEP event. We examine the 17 April 2021 SEP event that coincided with a long-lasting solar X-ray flare, a medium-speed CME, and multiple Type III radio bursts. The SEP proton datasets from this event were captured by the HET component of PSP/IS☉IS EPI-HI instrument; (note: HET-A is mounted with its axis 20° relative to the spacecraft's Z-axis, which is nominally pointed towards the Sun, Cohen et al. 2021b). This SEP event was observed by several spacecraft at varying solar distance, at which PSP was located at 0.42 AU. A more thorough description of the event can be found in Dresing et al. (2023). An overview plot of HETA data for the 17 April 2021 event is show in Figure 3 along with 1-minute magnetic field data from the PSP/FIELDS instrument.

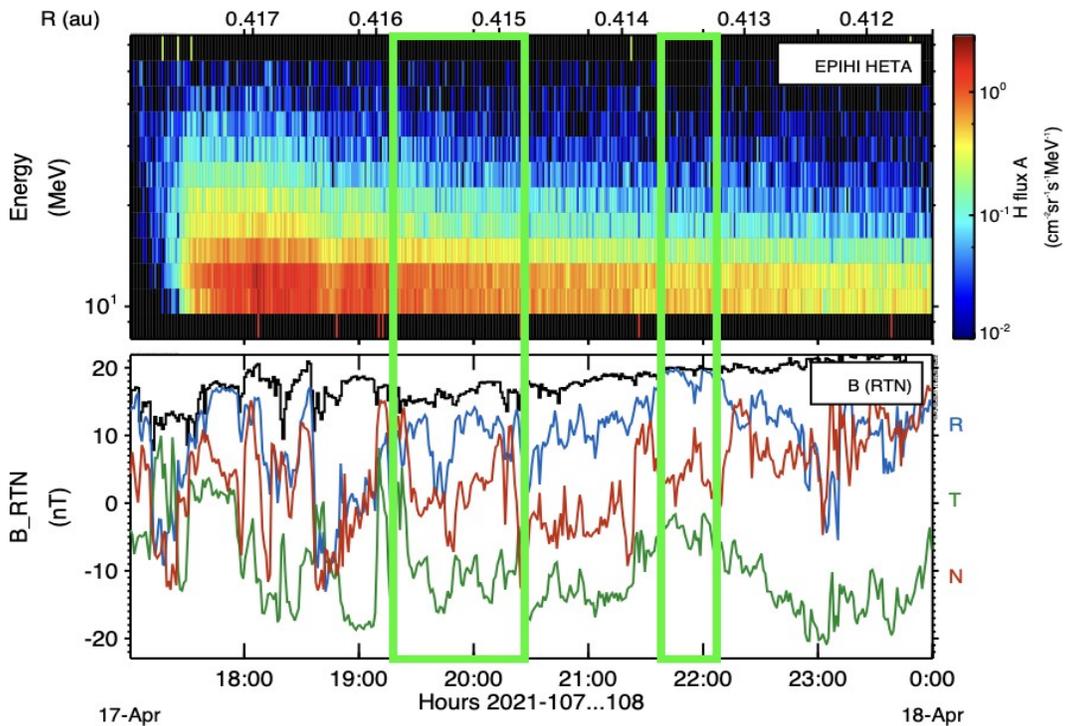

**Figure 3.** High energy proton observations from 2021-04-17 using the IS☉IS instrument's EPI-HI HETA band at 60 second cadence (upper panel) and magnetic field data from PSP's FIELDS instrument (lower panel). Green boxes outline the two magnetic flux tubes investigated in the paper. We remain on the first flux tube for 1 hour from 19:20 to 20:20. The second flux tube is shorter, from 21:30 to 22:10.

We analyze two different, non-overlapping time intervals from 17 April 2021. The first interval is from 19:20-20:20 and the second interval is from 21:30 to 22:10. We limit our selected energy range to 10-80 MeV, which are the middle bins of EPI-HI HETA, to avoid noise from the edge bins of the energy range. EPI-HI energy bins are distributed evenly in logarithmic space. For each minute, we extract the flux and its uncertainty values and plot the flux versus energy spectrum in a log-log scale. The raw flux from these intervals is plotted in Figure 3 and the time intervals we evaluate are denoted by the green boxes.

### 4.2. Methodology

We perform a linear fit to the flux spectrum using orthogonal distance regression to extract the (negative) slope and *y*-intercept. This regression method accounts for uncertainties both in dependent and independent variables, although we omit uncertainties associated with the detection location within each finite width energy bin. The spectral index is defined by the absolute value of the slope, which coincides with the thermodynamic kappa value $\kappa$. Figure 4 shows the linear fit for a single flux spectrum, constructed per 1-minute sampling in a one-hour observation window, as described in Section 4.1.

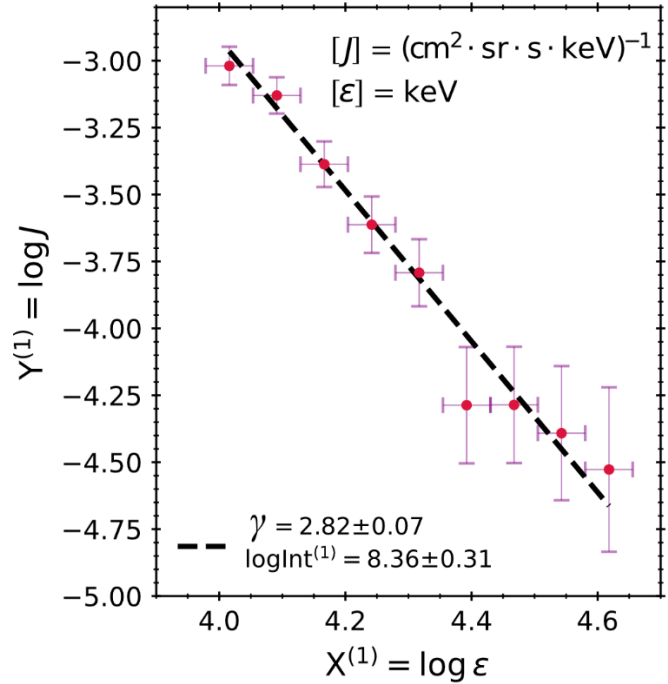

**Figure 4.** Plot of the energy-flux spectrum from a single 60-second observation with EPI-HI HET A, at 19:20 on 17 April 2023. Flux and Energy are plotted in log-log space together with their linear fit (black). Horizontal error bars represent the maximum and minimum energy for each energy bin (where all energy bins are spaced equally in logarithmic space). Flux uncertainties are derived from the HET A statistical uncertainties. The absolute value of the slope of the fitted line, i.e., the spectral index of the energy-flux spectrum, equals the value of the thermodynamic kappa.

### 4.3. Results

We perform the fitting routine for every minute within our observation window and collect the various values of spectral index $X^{(2)}$ and modified intercept, $Y^{(2)}$. (Note that values of thermodynamic kappa, corresponding to spectral index $\gamma \leq 1.6$, had no contribution to the fit to extract the temperature and density, because they were near the functional breaking point of $A(\gamma)$.) After collecting the aggregated pair values of $(X^{(2)}, Y^{(2)})$, we plot them and perform the linear fitting according to Eq.(14) (Section 3.1). We also perform a statistical test (Section 3.4) for detecting potential transition and other nonstationary features.

For the two examined time intervals, we derive the values of temperature *T* and density *n* of SEP protons from the linear fitting of the aggregate pair values of $(X^{(2)}, Y^{(2)})$. We find that as PSP passes through the CME, the SEP density of the second interval has decreased compared to the density of the previous time interval, while there is also a small decrease in temperature. This positive correlation between the evolution of temperature and density gives insights for a sub-adiabatic polytropic index, similar to other space plasmas, i.e., (e.g., see: Dayeh & Livadiotis 2022; Livadiotis et al. 2024), although solar wind plasma near the sun has shown to have super-adiabatic indices (Nicolaou et al. 2020). The aggregated pair values of $(X^{(2)}, Y^{(2)})$ and their linear fit (Section 3.1) and weighted average (Section 3.4), are shown in Figures 5(a) and 5(b), for the two intervals 19:20-20:20 and 21:30-22:10, respectively. Table 2 summarizes the results.

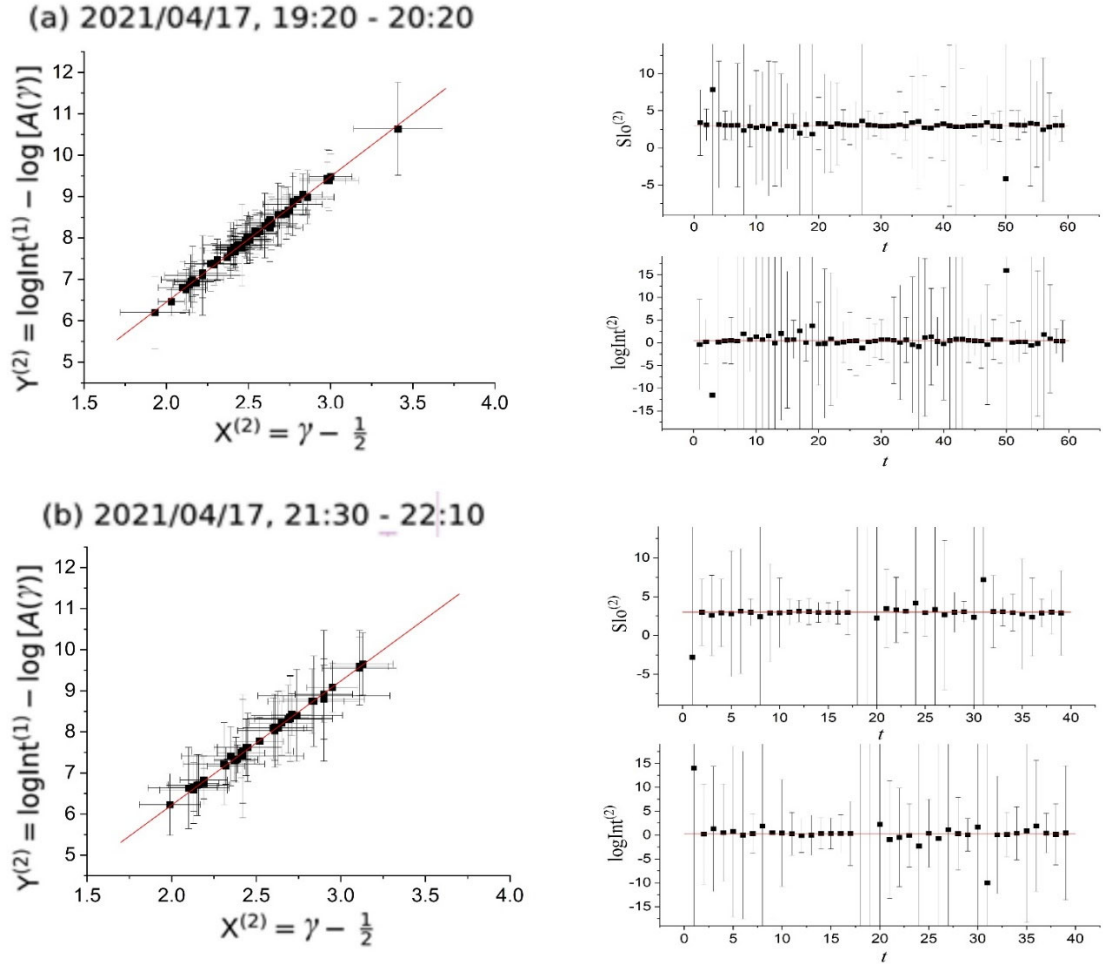

**Figure 5.** The aggregated pair values of $(X^{(2)}, Y^{(2)})$, defined in Eq.(11) as modified spectral index and intercept, determined for the intervals (a) 19:20 – 20:20 and (b) 21:30 – 22:10, both on the day 2021-04-17, are linearly related (with parameters noted by the line intercept logInt$^{(2)}$ and slope Slo$^{(2)}$). Their relationship is optimized either by the second level of linear fitting (left) (Section 3.1) or by the weighted average of the two-point estimated values (right) (Section 3.4 and Appendix A).

Table 2. Summary of the results

| | Period A - 19:20 – 20:20 | | Period B - 21:30 – 22:10 | |
|---|---|---|---|---|
| | logInt$^{(2)}$ | Slo$^{(2)}$ | logInt$^{(2)}$ | Slo$^{(2)}$ |
| **Fitting** | 0.381±0.076 | 3.035±0.031 | 0.174±0.070 | 3.023±0.028 |
| **Averaging** | 0.399±0.067 | 3.019±0.027 | 0.218±0.087 | 3.005±0.035 |
| $t_{est}$ | 0.173 | 0.408 | 0.398 | 0.408 |
| **Weighted Mean** | 0.391±0.050 | 3.026±0.020 | 0.191±0.055 | 3.015±0.022 |
| **Thermodynamics** | $n = (6.2±0.7)·10^{-7}$ cm$^{-3}$ | $T = 1060 ± 50$ keV | $n = (3.9±0.5)·10^{-7}$ cm$^{-3}$ | $T = 1040 ± 50$ keV |

(Note: All *t*-values result to p-values close to 1).

## 5. Discussion and conclusions

The paper has developed a novel kappa-tail fitting technique for analyzing observations of power-law tails of energy-flux spectra and connecting them with the theoretical modeling of kappa distributions. These distributions have their physical origin within the framework of thermodynamics, and thus, fitting a kappa distribution function to the energy-flux spectrum leads to the derivation of key thermodynamic parameters, such as, the temperature, density, and kappa, and consequently, characterize the thermodynamic state of the examined space plasma or particle population.

In general, the methodology of determining the thermodynamics of space particles involves employing the kappa distribution function, which has been widely utilized in low-energy space plasmas, such as the solar wind, to describe particle velocity distributions and the corresponding energy-flux spectrum throughout the heliosphere. The temperature parameter derived from the kappa distribution reflects the characteristic thermal energy of the particle population, while the density parameter represents the number density of particles within the system. Additionally, the kappa parameter itself serves as a measure of the degree of departure from a Maxwellian distribution, providing insights into the entropy of the particle ensemble; (the concept of entropy defect comprises the formulation of entropy and how this is parameterized by kappa, e.g., Livadiotis & McComas 2021; 2023a;b; 2024).

The kappa-tail fitting technique was shown in detail. In particular, we (i) developed the associated mathematical formulation, (ii) proved that the technique is decisive on whether the observed power-law is associated with kappa distributions; and (iii) provided a validation of the technique using pseudo-observations of typical input plasma parameters. The paper showed how the technique can be used for determining whether the tail is associated with kappa distributions, and if true, for deriving the temperature, density, and kappa of the examined space particles, here, SEP protons. The technique is based on two levels of linear fits, (1) the linear fit of the energy ($\varepsilon$) – flux ($J$) spectrum on log-log scales, that is, $X^{(1)} = \log(\varepsilon)$ and $Y^{(1)} = \log(J)$, for the estimation of the spectral index and intercept, and (2) the linear fit between the spectral index $X^{(2)}$ and modified intercept $Y^{(2)}$, where the new slope and intercept are respectively related to the temperature and density of the examined plasma particles, here, the energetic protons. A variation of the technique was developed, in the case where a transition between stationary states characterizes the examined time series interval.

As a case study, we applied the kappa-tail fitting technique to determine the thermodynamics of SEPs, observed by SPS/IS☉IS instrument. This proof-of-concept has related the kappa distribution function modeling with the SEP energy-flux spectrum for the two intervals in a SEP event, observed by PSP on April 17, 2021. Specifically, we (i) explained the data product from PSP and the application of the technique with which we assess thermodynamic properties of SEP events in PSP's observations; (ii) applied this technique to two separate intervals, and (iii) characterized the bulk temperature and density of high energy protons. In particular, the results are summarized: (1) the SEP density and temperature were found to decrease between the two different time intervals through the examined CME; on average, the results were $T \sim 1$ MeV and $n \sim 5 \cdot 10^{-7}$ cm$^{-3}$; (2) the observed positive correlation between density and temperature indicates a sub-adiabatic polytropic process; and (3) no transition has been detected in the examined intervals, but rather a single stationary state (described by the formulation of kappa distributions). Also, note that the thermodynamic kappa is well correlated to magnetic features captured by the FIELDS instrument, and thus, kappa can be further related to the magnetic flux tube crossing PSP.

This is the first study to systematically focus on the thermodynamics of SEPs. This paper comes along with the expanded study of Cuesta et al. (2024), which uses the developed technique and the kappa distribution formulation to model the tail of a collection of observed SEP energy-flux spectra. Further improvements of the technique may involve the following: (i) use the nonlinear fitting model of Eq.(8a) to describe the first level of fit; (ii) allow the dimensionality to be different than $D=3$ and given by the effective degrees of freedom (e.g., Livadiotis & McComas 2023b; Livadiotis et al. 2024); and (iii) optimize the size and statistical features of the running window that produces the pair values of $(X^{(2)}, Y^{(2)})$ for the 2$^{nd}$ level of fitting.

This research aims, in general, to enhance our understanding of the thermodynamic properties of energetic protons and other SEPs from the solar corona out through the heliosphere and in association with bulk solar wind bulk protons and interplanetary magnetic field. Also, this research may ultimately contribute to advancing space weather forecasting capabilities, aiding in the mitigation of potential risks posed by SEPs to both space missions and human exploration beyond Earth's orbit. Furthermore, the derived thermodynamic parameters serve as essential inputs for theoretical models aimed at elucidating the underlying physical mechanisms governing SEP acceleration and transport processes. The showcased technique allows the space physics community to explore the thermodynamic characteristics of any particle population, utilizing measurements of energetic particles from PSP/IS☉IS and many other missions.

**Acknowledgements**

We thank the IS⊙IS team and everyone that made the PSP mission possible. The IS⊙IS data and visualization tools are available to the community at https://spacephysics.princeton.edu/missions-instruments/PSP. PSP was designed, built, and is operated by the Johns Hopkins Applied Physics Laboratory as part of NASA's Living with a Star (LWS) program (contract NNN06AA01C). This research was partially funded by PSP GI grant 80NSSC21K1767.

**Appendix A. Variation of the technique**

Here, we describe the variation of the technique, that examines time intervals with potential transitions between stationary states. Let the number of observations with uncertainties, $\{y_i \pm \sigma_{yi}\}_{i=1}^{N}$, described by the statistical model $f(x;\{p_k\}_{k=1}^{M})$ that depends on $M$ independent parameters, $\{p_k\}_{k=1}^{M} = p_1, p_2, ..., p_M$. Instead of fitting directly the model to the whole $N$-point dataset, the method uses subsets of $M$-point data. Using each of these subsets, we form $M$-equations $\{y_j = f(x_j;\{p_k\}_{k=1}^{M})\}_{j=1}^{M}$, which we can solve to find the $M$ unknown parameters, $p_k = p_k(\{x_j, y_j\}_{j=1}^{M})$. The respective propagation errors are also estimated, $\delta p_k = \sqrt{\sum_{j=1}^{M}(\partial p_k/\partial y_j)^2 \sigma_{yj}^2}$, or $\delta p_k = \sqrt{\sum_{j=1}^{M}[(\partial p_k/\partial y_j)^2 \sigma_{yj}^2 + (\partial p_k/\partial x_j)^2 \sigma_{xj}^2]}$, in the case where both independent variables have uncertainties, $\{x_i \pm \sigma_{xi}\}_{i=1}^{N}$ and $\{y_i \pm \sigma_{yi}\}_{i=1}^{N}$. Each $M$-point data subset provides a calculated value for each parameter and its propagation error.

We can construct a number of $\binom{N}{M}$ possible combinations of $M$-point data subsets, each suitable for the method to estimate the parameters $\{p_k \pm \delta p_k\}_{k=1}^{M}$. Nevertheless, when collecting the $M$-point data subsets, we examine whether or not it is physically meaningful to keep the order of the data index $i$; e.g., when represents the observation time, then, the $M$-point subsets are chosen to be sequential to preserve the physical order. Then, a dataset of $N$ observations can be used for construction of a moving window of $M$-point data subset, leading to the construction of $N-M+1$ values of each parameter. However, if there is no any desired $i$-index direction, then, any of the possible $\binom{N}{M}$ combinations are suitable for the method.

Next, we compare all the estimated values of each parameter. Let $s$ note the total number of estimated values of each parameter, i.e., $\{(p_k)_1, (p_k)_2, \cdots, (p_k)_s\}_{k=1}^{M} = \{\{(p_k)_n\}_{n=1}^{s}\}_{k=1}^{M}$. This number $s$ depends on the sampling of $M$-point subsets: if there is a moving window along data index $i$, then $s = N-M+1$; however, if we collect combinations of $M$-point subsets, then, $s \leq \binom{N}{M}$, with the equality holding when all the possible combinations are collected. In any case, all the $s$ estimated values of each of the $M$ parameters, are physically equivalent. Namely, they are associated with the statistical hypothesis that are equal. Thus, they can be statistically described by a trivial fitting model, that is, a constant, $f(x;p) = p$, and the optimal fitting value of this constant is the average of the involved parameter values:

$$\chi_k^2 \equiv \sum_{n=1}^{s} \left( \frac{(p_k)_n - \overline{p}_k}{(\delta p_k)_n} \right)^2 , \qquad (A1)$$

where the optimal value of the respective parameter $p_k$ is given by the weighted mean

$$\overline{p}_k = \frac{\sum_{n=1}^{s} (\delta p_k)_n^{-2} (p_k)_n}{\sum_{n=1}^{s} (\delta p_k)_n^{-2}} . \qquad (A2)$$

The respective propagation and fitting errors are

$$\delta \overline{p}_{k,\text{prop}} = \frac{1}{\sqrt{\sum_{n=1}^{s} (\delta p_k)_n^{-2}}} , \quad \delta \overline{p}_{k,\text{fit}} = \sqrt{\left(\overline{p_k^2} - \overline{p}_k^2\right) / N_e} , \qquad (A3)$$

which includes the weighted mean square and the effective degrees of freedom:

$$\overline{p_k^2} = \frac{\sum_{n=1}^{s} (\delta p_k)_n^{-2} (p_k)_n^2}{\sum_{n=1}^{s} (\delta p_k)_n^{-2}} , \quad N_e = \left[ \sum_{n=1}^{s} (\delta p_k)_n^{-2} \right]^2 / \sum_{n=1}^{s} (\delta p_k)_n^{-4} - 1 , \qquad (A4)$$

(see: Melissinos 1966, Bevington 1969; Livadiotis 2007, 2014b, 2018c; Zirnstein et al. 2016; Livadiotis et al. 2022). (Note: The two types of errors are the same in the case where the reduced chi-square is ~1, and thus, the fitting is characterized as good. Otherwise, there is a potential overestimation in the values of the propagation errors, and the fitting errors are those that should be used instead.)

In the application of the method on SEPs, we have $M=2$ parameters to estimate, the slope $\text{Slo}^{(1)}$ and intercept $\log \text{Int}^{(1)}$, or, the spectral index $\gamma$ and the modified intercept, $\log \text{Int}' \equiv \log \text{Int}^{(1)} - \log[A(\gamma)]$, from which we construct the pair of $\{X^{(2)} \equiv \gamma - \tfrac{1}{2}, Y^{(2)} \equiv \log \text{Int}'\}$. We select a number of $N$ sequential spectra, from which we estimate $N$ pair values, $\{X^{(2)}_i, Y^{(2)}_i\}_{i=1}^{N}$.

Using a moving window of two points along the whole interval of $N$ points, we construct a number of $N-1$ two-point subsets, each characterized by a value of the two parameters, slope $\text{Slo}^{(2)}$ and intercept $\log \text{Int}^{(2)}$. For simplicity, we rename $\{x_i \equiv X^{(2)}_i, y_i \equiv Y^{(2)}_i\}_{i=1}^{N}$, $a \equiv \log \text{Int}^{(2)}$ and $b \equiv \text{Slo}^{(2)}$. Then, for any two-point subset, we equate the data with linear model, $y_i = a + b \cdot x_i$ and $y_{i+1} = a + b \cdot x_{i+1}$, leading to

$$a = \frac{y_i x_{i+1} - y_{i+1} x_i}{x_{i+1} - x_i} , \quad b = \frac{y_{i+1} - y_i}{x_{i+1} - x_i} , \qquad (A5)$$

and

$$\delta a = \frac{\sqrt{x_{i+1}^2 \delta y_i^2 + x_i^2 \delta y_{i+1}^2 + b^2(x_{i+1}^2 \delta x_i^2 + x_i^2 \delta x_{i+1}^2)}}{|x_{i+1} - x_i|} \quad, \quad \delta b = \frac{\sqrt{\delta y_i^2 + \delta y_{i+1}^2 + b^2(\delta x_i^2 + \delta x_{i+1}^2)}}{|x_{i+1} - x_i|} \quad . \quad (A6)$$

Using these formulae, we derive the values of the parameters for all the *N*-1 two-point subsets, $\{\log \text{Int}^{(2)}_i, \text{Slo}^{(2)}_i\}_{i=1}^{N-1}$. Then, we proceed to the weighted average of the *N*-1 parameter values, $\overline{\log \text{Int}^{(2)}}$ and $\overline{\text{Slo}^{(2)}}$. These estimated averages, $\overline{\log \text{Int}^{(2)}}$ and $\overline{\text{Slo}^{(2)}}$, can be used in a statistical test (see main text, Section 3.4) to determine whether the examined interval is stationary or is characterized by transition between stationary states.

**Appendix B. Simulations of transitions between stationary states**

The method described in this paper provides a way to determine the density and temperature from the variations of the energetic particle intensity spectra. In general, there is an entire set of possibilities for the density that could generate a series of spectra. Based on the thermodynamic theory that this method is based upon, variations in density, temperature, and kappa are expected to be related by the polytropic relationship $n \propto T^\nu$, where the exponent $\nu$ measures the polytropic index; the connection between kappa and polytropic index is $\nu + \kappa = \frac{1}{2}$ (e.g., see: Livadiotis 2016; 2019; Livadiotis et al. 2022; note that in the main text, this relation is written as $\nu + \gamma = \frac{1}{2}$).

In practice, even if the linear relationship indicates that a kappa distribution is behind the observed spectra at a series of points, there may be a transition between these plasma parcels that is not consistent with this polytropic relationship in special circumstances. It is important to consider how the proposed technique is affected by such cases.

Consider a series of fictitious hot, dense plasma blobs with hardened energetic particle spectra encountered by a spacecraft. From the kappa polytropic behavior, we would expect an anticorrelation between density and temperature, but the plasma blobs contradict this expectation. Also, consider further microscopic fluctuations superposed over these macroscopic plasma blobs. One possibility is that they exhibit kappa polytropic behavior. Another possibility is that they exhibit behavior inconsistent with this due to nonideal behavior, e.g. higher-energy particles arrived causing a change in the spectral index without immediately changing the polytropic behavior.

A simulation of these scenarios is presented in Figure 6. The spectra are fitted alternatively with *M*=2 and *M*=*N*=31 points (following the notation in Appendix A). In the case where the background fluctuations follow the kappa polytropic behavior, the method gives stable results which accurately reflect the ground truth regardless of the window size. During nonideal periods, such as the boundaries of the plasma blobs, the output fluctuates wildly by orders of magnitude when two-point fitting is used (Appendix A). In contrast, the 31-point fitting forces the output to agree with the polytropic behavior expected for the given value of kappa (anticorrelation between density and temperature) while maintaining a smooth output. Hence, the fit using a large number of points can be misleading when there is nonideal behavior.

In the case where the nonideal behavior is gradual, as in the plasma blob boundaries, the median value of the two-point output accurately describes the density and temperature variations. On the other hand, when the background fluctuations are also nonideal, the median is significantly less reliable even using two-point fitting. Nevertheless, the two-point fitting is useful in this case as well because the range of values becomes extremely large compared to the median, as illustrated in the fourth panel. Thus, the deviation from the median using two-point fitting can quantify the reliability of the output. (Note: The median is associated with the taxicab norm, instead of the standard Euclidean norm; a property that makes it less sensitive to fluctuations; see: Livadiotis 2012.)

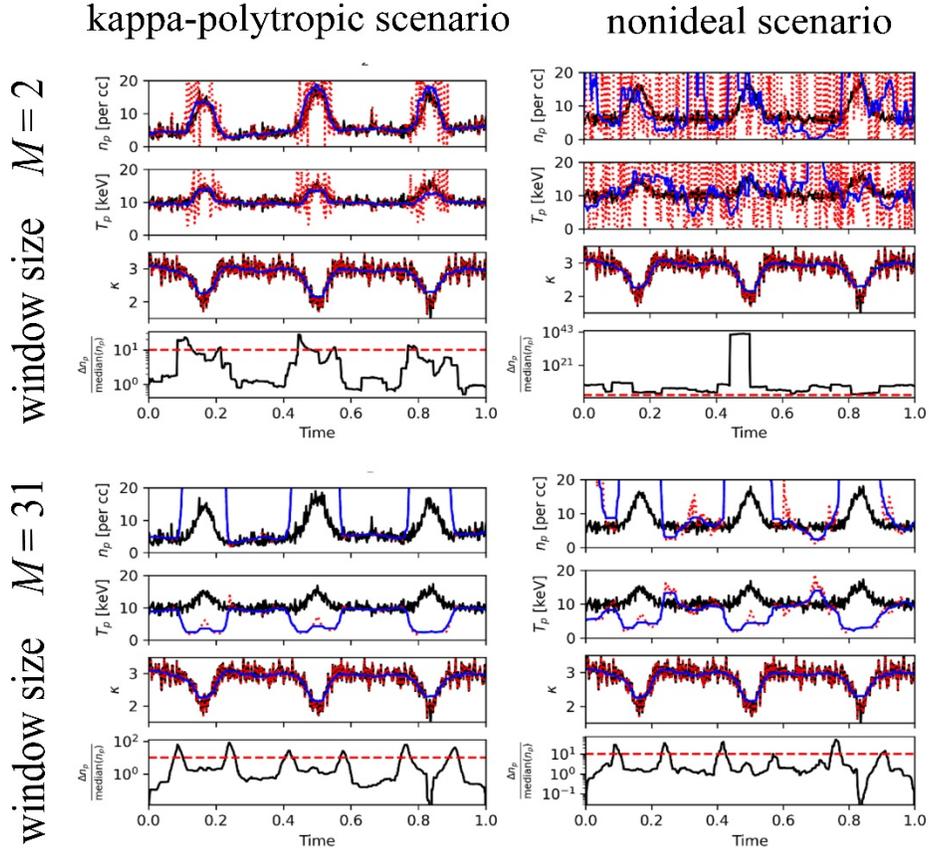

**Figure 6.** Simulations of hot, dense plasma blobs within a background of small-scale fluctuations. The fluctuations in kappa and temperature are generated using normally distributed random deviations from the mean. The variations in density are chosen according to $n \propto T^{\nu}$. The hot, dense plasma blobs are generated with an addition proportional to $\sin^{\chi}(\omega t)$, where the frequency $\omega$ and exponent $\chi$ are chosen so that the interval contains three sharp peaks in density and temperature and decreases in kappa (corresponding to a steeper power law tail). The black lines represent the ground truth. The dotted red lines represent the raw output of the technique while the blue line is a rolling median of the raw outputs.